# INVESTIGATION OF AN ALTERNATE MEANS OF WAKEFIELD SUPPRESSION IN THE MAIN LINACS OF CLIC[*]

V. F. Khan, R.M. Jones; The Cockcroft Institute, Daresbury, WA4 4AD, UK;
The University of Manchester, Manchester M13 9PL, UK.

## Abstract

We report on suppression of long-range wakefields in CLIC accelerating structures. Strong detuning and moderate damping is employed. In these initial design studies we focus on the CLIC_G structure and enforce a moderate Q of 300 and 500. We maintain a dipole bandwidth of approximately 1 GHz as specified from breakdown constraints in a modified structure, CLIC_DDS. The circuit model, which facilitates a rapid design of manifolds slot-coupled to the main accelerating cells, is described.

## INTRODUCTION

The CLIC design for a lepton machine aims at 3 TeV centre of mass collisions through acceleration with normal conducting (NC) linacs operating at a gradient of 100 MV/m. The ultra-relativistic beam excites an electromagnetic (e.m.) field which has both intra-bunch (short-range) and inter-bunch (long-range) components, represented as wakefields. This wakefield has a longitudinal component, which effects the energy spread of the bunches and, a transverse component which dilutes the beam emittance and can give rise to a BBU [1] instability. In this work we investigate suppression of the transverse component of the long-range wakefield.

These wakefields, in the present baseline CLIC design, [2] are suppressed by waveguides which couple out the fields from each accelerating cell. This results in strong damping (Q~10) and entails dielectric damping materials being closely located to the cells. We present an alternate design which entails strong detuning of the cell frequencies and moderate damping (Q ~ 300-500). The latter is reflected through four manifolds running along the walls of the accelerator. The broad principles of the design are similar to that used in the NLC/GLC [3]. However, the requirements for CLIC on wakefield suppression are more stringent, as the bunches are separated from one another by ~ 0.5 ns, compared to 1.4 ns in the NLC/GLC [4]. Our original design [5], with an enforced first band dipole bandwidth of ~3 GHz, enabled the wakefields to be adequately suppressed for all trailing bunches. However, electrical breakdown constraints imposed, due to ambitious accelerating field gradients sought, contain the allowable iris range and hence the bandwidth has been restricted to ~ 1 GHz. Here we present initial results based on these restrictions.

## UNCOUPLED CELL DISTRIBUTION

The present CLIC baseline design for each accelerating structure consists of 24 cells with an iris range of 3.15 mm to 2.35 mm. Tapering the irises within the structure ensures that a range of modes are excited which do not add coherently. We enforce an erf distribution of the dependence of the iris radius with cell number. The mode density dn/df, characteristic kick factor weighted density function Kdn/df, together with an indication of mode location, for a structure effectively consisting of four times the number of cells, is illustrated in Fig. 1.

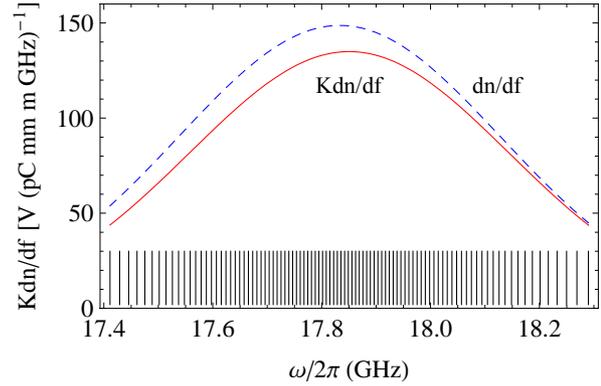

Figure 1: Kick factor weighted density function Kdn/df (red curve), density function (blue-dashed) and mode separation (black lines).

The Gaussian distribution used in this design is not necessarily the optimum distribution as other functional forms may give rise to a more rapid decay of the wakefield [6]. The Fourier transform of the N-cell Gaussian Kdn/df distribution provides an indication of the fall-off of the wakefield [5]. However, as the tails of the Gaussian distribution are truncated, a more accurate calculation of the envelope of the uncoupled wakefield is given by [5]:

$$W_t = 2\overline{K} e^{-2(\pi\sigma t)^2} |\chi(t, \Delta f)| \quad (1)$$

where $\overline{K}$ is the average kick factor, the $\chi$ factor is defined in [5] and $\Delta f$ represents the dipole bandwidth of:

$$K\frac{dn}{df} = N\frac{\overline{K}}{\sqrt{2\pi}\sigma} e^{-(f-f_0)^2/2\sigma^2} \quad (2)$$

We chose a bandwidth of 3σ or ~0.05$f_0$. In order to calculate the long-range behaviour of the wake function we utilise a circuit model which includes TE-TM dipole modes coupled via TE modes to a manifold through slots cut into each cell. This circuit model includes a double chain of L-C circuits capacitively coupled to a transmission line [3]. The wakefield is calculated by a spectral method which requires 8 parameters and the corresponding kick factors to be obtained for each cell of the accelerating structure. We parameterise the structure by choosing 7 fiducial cells and the parameters of the other cells are obtained by cubic spline interpolation. This procedure allows an efficient determination of the wakefield. The parameters are subsequently made functionally dependent on the synchronous frequencies of each cell. This facilitates a rapid evaluation of the wakefield for new distributions. The parameters of each cell are obtained through solving 8 coupled non-linear equations, resulting from the circuit equations, for each fiducial cell subjected to infinite periodic boundary conditions. The circuit equation describing a dipole mode of frequency f coupled to the manifold at a phase advance per cell of $\psi$ is given by [3]:

$$\left\{\left[(1+\eta\cos\psi)/f_0^2 + \Gamma^2/\alpha/(F^2 - f^2) - f^{-2}\right]\right.$$
$$\left[(1-\hat{\eta}\cos\psi)/\hat{f}_0^2 - f^{-2}\right] - \eta_x^2/(f_0\hat{f}_0)\sin^2\psi\right\}$$
$$[\cos\psi - \cos\varphi] = $$
$$\Gamma^2 F^2/(F^2 - f^2)(\pi L/c)^2\left[(1-\hat{\eta}\cos\psi)\hat{f}_0^{-2} - \hat{f}^{-2}\right]\text{sinc}\varphi$$
(3)

where all parameters are described in [3]. The corresponding dispersion curves are illustrated in Fig. 2.

---

[*] Research leading to these results has received funding from European commission under the FP7 research infrastructure grant no. 227579.

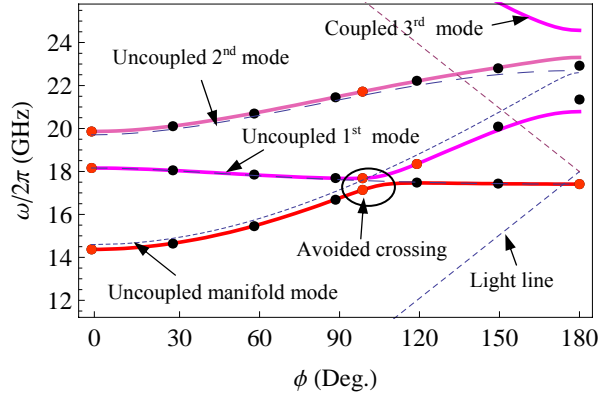

Figure 2: Brillouin diagram for cell 1 of CLIC_DDS. Points represent simulations and the solid (long-dashed) curves represent the coupled (uncoupled) dipole modes. The short-dashed curves represent the uncoupled manifold mode.

All points have been obtained with HFSS v11 operating in the eigenmode module. The red points are used in solving the coupled non-linear equation and the curves are required to pass through these points. The additional (black) points provide an indication as to the quality of the fits and are not used in the parameter determination. The field distribution of the lower 0 and the lower $\pi$ mode are illustrated in Fig. 3.

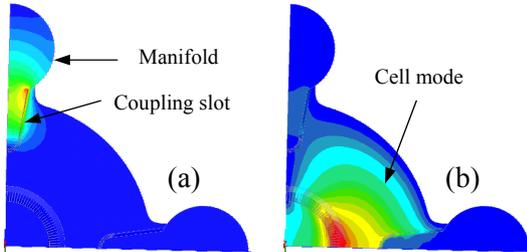

Figure 3: Electric field in a CLIC_DDS cell with quarter symmetry at 0 phase (a) and $\pi$ phase (b), corresponding to $\omega/2\pi$ = 14.37 GHz and 17.41 GHz respectively

The dipole mode is excited at the point where the light line intersects the curves, at the synchronous point. This mode is in the vicinity of a $\pi$ phase advance and Fig. 3(b) indicates the expected dipole mode field; in this case it is confined to the region of the cell. The mode subsequently travels down the accelerating structure and couples out to the manifold through slots downstream. Once the parameters of all 7 cells have been obtained in this manner, others are obtained by interpolation. This facilitates the coupled wakefield to be determined through the spectral method

## SPECTRAL CALCULATION OF THE WAKE FUNCTION

The matrix describing the currents excited in each of the cells, modelled by coupled L-C circuits [7], is:

$$\begin{pmatrix} \hat{H} - \lambda U & H_x^t & 0 \\ H_x & H - \lambda U & -G \\ 0 & G & -R \end{pmatrix} \begin{pmatrix} \hat{a} \\ a \\ A \end{pmatrix} = \begin{pmatrix} \lambda B \\ 0 \\ 0 \end{pmatrix}$$
(4)

or in a condensed form:
$$(\overline{H} - \lambda U)\overline{a} = \lambda \overline{B}$$
(5)

Here, $f = \omega/2\pi = 1/\sqrt{\lambda}$ is the frequency, H ($\hat{H}$) describes TE (TM) modes, $H_x$ is the cross coupling matrix (between TE and TM modes). G and R are manifold coupling matrices and U is the identity matrix. Each sub-matrix within the matrix described by Eq. (4) is of dimension N x N (N = 96 for the 4-fold interleaved CLIC structure). The eigenvalues ($\lambda$) and eigenmodes of the above matrix allow the coupled kick factors and coupled mode frequencies to be determined. The envelope of the transverse wakefield can, in principle, be determined as a sum over modes [5]. However, as the coupling is particularly large, the shift in the coupled frequency compared to uncoupled values changes the character of the modes. For this reason we use the spectral function method [3], specially developed for this purpose. The spectral function $S(\omega)$ is given by $4\text{Im}\{Z(\omega)\}$, where:

$$Z(\omega) = \frac{1}{2\pi^2} \sum_{n,m}^{N} \sqrt{K_s^n K_s^m \omega_s^n \omega_s^m} \exp\left[(j\omega L/c)f(n-m)\right] \tilde{H}_{nm}$$
(6)

Here the 3N x 3N matrix, $\tilde{H}$, is given by:
$$\tilde{H} = \overline{H}(U - \lambda^{-1}\overline{H})^{-1}$$
(7)

L refers to the length of each accelerating cell and $K_s^n$ is the single-cell kick factor evaluated at the synchronous frequency $\omega_s/2\pi$ for mode n. All the other parameters are as described in [3]. The corresponding envelope of the transverse wakefield in the time domain is:

$$W_t(t) = \frac{\theta(t)}{2\pi}\left|\int_0^\infty S(\omega)\exp(j\omega t)\,d\omega\right|$$
(8)

Here $\theta(t)$ is the Heaviside step function. We apply this technique to a modified parameter set based on CLIC_G to populate all matrices in Eq. (4). In order to provide adequate sampling of the uncoupled $2K dn/df$ distribution (Fig. 1) the mode frequencies are interleaved by utilising 4xN cells in the spectral function calculation. In this manner, the dipole frequencies of cells in neighbouring

structures are interleaved. The corresponding spectral function is illustrated in Fig. 4. No attempt has been

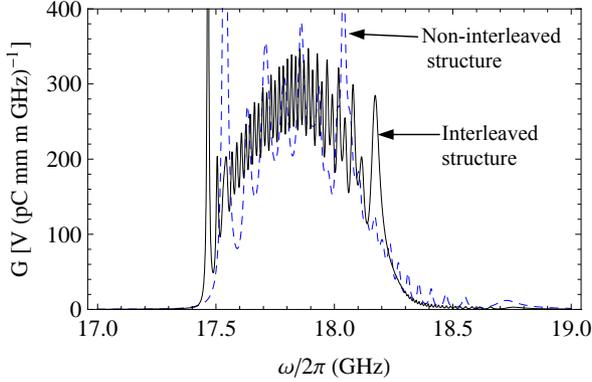

Figure 4: Spectral function for non-interleaved (blue-dashed) and four-fold interleaved (black) CLIC_DDS.

Made to optimally position the location of these interleaved frequencies. The modes with large Q values are not well damped. A careful optimisation achieved through non-linear positioning of the interleaved frequencies will ameliorate this problem. The modal Qs, calculated from Lorentzian fits to the spectral function, (Fig. 4) are displayed in Fig. 5. The average Q is ~ 500.

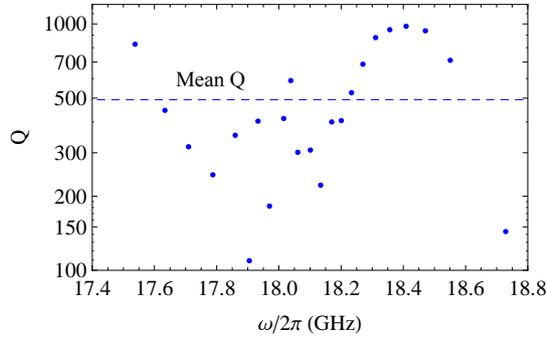

Figure 5: Lowest band dipole Qs for a single CLIC_DDS (non-interleaved) structure.

However, there are some significantly large Qs which would benefit from non-linear optimisation. The transverse wake function corresponding to that given in Eq. (8) is displayed in Fig. 6. Here the manifold coupling and related modal Qs are explicitly taken in into account.

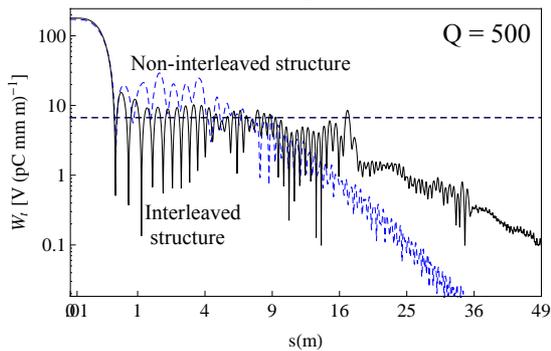

Figure 6: Envelope of wake function for CLIC_DDS. Non-interleaved (blue-dashed) and interleaved (black) are calculated. The average modal Q ~ 500.

Interleaving has the beneficial effect of reducing the average of the envelope wake in the first ~ 4 m by a factor of ~ 2. However, the wake is still unacceptable from a beam dynamics perspective. In order to further understand the effect of enhancing the mode Qs we made an *aposteri* application of an enforced modal Q ~ 300 to all modes. The resulting wake function is displayed in Fig 7. This indicates that provided optimisation of the mode damping results in a mode Q of 300, the wake function will be appropriately damped apart from the first three trailing bunches. The wake experienced by the first three bunches can be damped by relaxing the 1 GHz dipole mode bandwidth and by positioning them at the location of the zero crossing in the wake function. The latter will require an exhaustive set of beam dynamics simulations in order to account for realistic fabrication tolerances [7].

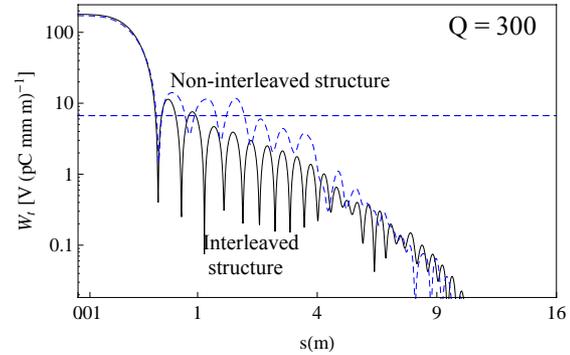

Figure 7: Envelope of wake function for CLIC_DDS with a Q of 300 imposed over all modes showing non-interleaved (blue-dashed) and interleaved (black).

Application of the spectral function method enables structures to be designed to suppress the wake function in the CLIC main linacs. The initial design presented here, is modified by a design to reduce the maximum pulse temperature rise on the wall of the cavity and to remotely locate the HOM loads. Four-fold interleaving of the erf distribution of frequencies of successive structures has been employed. This design can benefit from additional optimisation. Further designs are anticipated utilising different Kdn/df distributions and will benefit from a modestly increased bandwidth. These modifications are under active consideration. Furthermore, we are now in a position to perform an optimisation of the manifold-to-cell coupling and on means to further reduce the magnetic field in the vicinity of the coupling slots where it is maximum with respect to the whole structure.